\documentclass[aps,prl,twocolumn,showpacs,groupedaddress]{revtex4}
\usepackage{graphicx}
\begin{document}
%\preprint{}


\title{Evidence for a full energy gap for nickel-pnictide LaNiAsO$_{1-x}$F$_x$ superconductors by $^{75}$As nuclear quadrupole resonance}

\author{T. Tabuchi $^1$}
\author{Z. Li $^1$}
\author{G.F. Chen $^2$}
\author{S. Kawasaki $^1$}
\author{J.L. Luo $^3$}
\author{N.L. Wang $^3$}
\author{Guo-qing Zheng $^{1,3,*}$}

\affiliation{$^1$Department of Physics, Okayama University, Okayama 700-8530, Japan} 
\affiliation{$^2$Department of Physics, Renmin University of China, Beijing 100872, China }
\affiliation{$^3$Institute of Physics and Beijing National Laboratory for Condensed Matter Physics, Chinese Academy of Sciences, Beijing 100190, China }


%\author{}
%\email[zheng@psun.phys.okayama-u.ac.jp]{Your e-mail address}
%\thanks{}
%\affiliation{}


\date{\today}

\begin{abstract}
We report  systematic $^{75}$As-NQR and $^{139}$La-NMR studies on nickel-pnictide superconductors LaNiAsO$_{1-x}$F$_x$ ($x$=0, 0.06, 0.10 and 0.12).   The spin lattice relaxation rate $1/T_1$ decreases below $T_{\rm c}$ with a well-defined coherence peak and follows an exponential decay at  low temperatures. This result indicates that the superconducting gap is fully opened, and is strikingly different from that observed in iron-pnictide analogs. In the normal state,  $1/T_1T$ is constant in the temperature range $T_{\rm c}\sim$4 K$\leq T \leq$10 K for all compounds and up to $T$=250 K for $x$=0 and 0.06, which indicates weak electron correlations and is also different from the iron analog. We argue that the differences between the iron and nickel pnictides arise from the different electronic band structure. Our results highlight the  importance of the peculiar Fermi-surface topology in iron-pnictides.  
\end{abstract}

\pacs{74.25.-q, 74.25.Nf, 74.90.+n}
%\keywords{}

\maketitle


%\section{}\label{}
Superconductivity in ReFeAsO$_{1-x}$F$_x$ (Re: rare earth element) at the  transition temperature up to $T_{\rm c}$ = 55 K \cite{Kamihara,Ren3} has received considerable attention. 
These compounds have a ZrCuSiAs type structure (P4/nmm)  in which FeAs forms a two-dimensional network similar to the CuO$_2$ plane in the case of cuprate high-$T_{\rm c}$ superconductors. By replacing O with F, electrons are doped and superconductivity emerges. The Fermi surface consists of  two hole-pockets centered at (0, 0) ($\Gamma$ point) of the unfolded Brillouin zone, and two electron-pockets around ($\pi$, 0)  \cite{Singh}. Nuclear magnetic resonance (NMR) measurement in PrFeAsO$_{0.89}$F$_{0.11}$ first suggested that there are multiple superconducting gaps    \cite{Matano}. Angle resolved photoemission spectroscopy (ARPES) has directly observed the gaps on different Fermi surfaces \cite{Ding}. 

However, the symmetry  of the superconducting gap  and the mechanism of the superconductivity remains unclear. 
The spin lattice relaxation rate $1/T_1$ shows neither a coherence peak just below $T_c$, nor  an exponential decay at low temperature ($T$) \cite{Matano,Ishida,Grafe}  expected for  conventional fully-gapped superconductors. The result was interpreted as indicative of  nodes in the gap function. On the other hand, tunneling \cite{Chien} and ARPES  \cite{Ding}  measurements  have suggested a full gap. The  penetration depth measurements by different groups \cite{Matsuda,Ames} have led to different conclusion on whether there are nodes in the gap function or not.

 In the normal state, antiferromagnetic spin fluctuations with wave vector $Q$=($\pi$, 0) is expected due to the nesting between the electron-like Fermi surface and the hole-like Fermi surface \cite{Mazin,Kuroki,Chubukov,Ogata}. Indeed, signatures of such spin fluctuations are seen in  LaFeAsO$_{0.92}$F$_{0.08}$ \cite{Kawasaki},  Ba$_{1-x}$K$_{x}$Fe$_2$As$_2$ \cite{Matano2,Fukazawa,Yashima} and Fe$_{1-x}$Se \cite{Imai-FeSe} where $1/T_1T$ increases with decreasing temperature. It has been proposed that such  spin fluctuations may promote superconductivity with $s^{\pm}$-wave gap
  that changes sign on different Fermi surfaces \cite{Mazin,Kuroki,WangF}. In such case, the scatterings between different Fermi surfaces may reduce the coherence peak in the $T$-dependence of $1/T_1$ \cite{Chubukov,Parker,Hu,Bang,Nagai}, thereby reconcile the discrepancy between NMR and tunneling/ARPES measurements. However,  experimental verification of such exotic superconducting state remains to be carried out.

%It is noted that the inter-band magnetic scattering with the wave vector $q$=($\pi$,0) is responsible for the sign change $s$-wave gap. 

%The interband scattering is proposed to be also resposible for the magnetic correlations   

Meanwhile, the nickel analog of  LaFeAsO$_{1-x}$F$_x$, namely LaNiAsO$_{1-x}$F$_x$, are superconducting, but with a lower $T_c\sim$ 4 K \cite{Li}. LDA calculation has revealed a striking difference in Fermi surface compared to the Fe-analog.  Namely, there are no hole pockets around the (0, 0)  point  for LaNiAsO$_{1-x}$F$_x$ \cite{Xu}. Since interband scattering between the electron and hole pockets with the wave vector $Q$=($\pi$, 0) is proposed to be responsible for both the superconductivity and the normal-state properties for LaFeAsO$_{1-x}$F$_x$,  LaNiAsO$_{1-x}$F$_x$ without such Fermi surface nesting  is an ideal compound to compare with and to obtain insight into LaFeAsO$_{1-x}$F$_x$. In particular, identifying the superconducting gap symmetry and the nature of electron correlations, if any, in LaNiAsO$_{1-x}$F$_x$ will help test the theoretical proposals and understand the mechanisms of high-$T_c$ superconductivity in iron-pnictides.

In this Letter, we report the first microscopic measurement on  LaNiAsO$_{1-x}$F$_x$ ($x$=0, 0.06, 0.10 and 0.12) using the $^{75}$As nuclear quadrupole resonance (NQR)   and $^{139}$La-NMR techniques. 
The spin-lattice relaxation rate, $1/T_1$, shows a well-defined coherence peak just below $T_c$ and decays exponentially with further decreasing $T$, indicating a fully opened superconducting gap on the entire Fermi surface. 
This is the first clear NMR/NQR evidence for a full gap in the pnictide superconductors.
In the normal state above $T_c$, no antiferromagnetic spin fluctuations were observed. These features are  striking different from LaFeAsO$_{1-x}$F$_x$ where high-$T_c$, unusual superconductivity emerges with   moderate strength of antiferromagnetic spin fluctuation as a background. The difference is understood by the different topology of the Fermi surface of the two classes of materials. Our results highlight the importance of the Fermi surface topology in Fe-pnictides and shed new lights on  the mechanism of  superconductivity and electron correlations there.

The polycrystalline samples   of LaNiAsO$_{1-x}$F$_x$ were  synthesized by the solid-state reaction method using NiO, Ni, As, La, and LaF$_3$ as starting materials \cite{Li}. LaAs was prepared by reacting La chips and As pieces at 500 $^0$C for 15 h and then at 850 $^0$C for 2 h. The raw materials were
thoroughly ground and pressed into pellets and then annealed at
1150 $^0$C for 50 h. For NQR measurements, the pellets were ground into powder.
$T_c$ was determined by DC susceptibility measurement using a SQUID and by AC susceptibility measurement using the in-situ NQR coil. The results by the two methods agree well.  The NQR spectra were taken  by changing the frequency point by point, while the NMR spectra were taken by sweeping the magnetic field at a fix frequency. 
The $1/T_1$ was measured by using a single saturation pulse.  Measurements below $T$=1.3 K was conducted by using a $^3$He-$^4$He dilution refrigrator.

\begin{figure}[h]
\includegraphics[width=7.0cm]{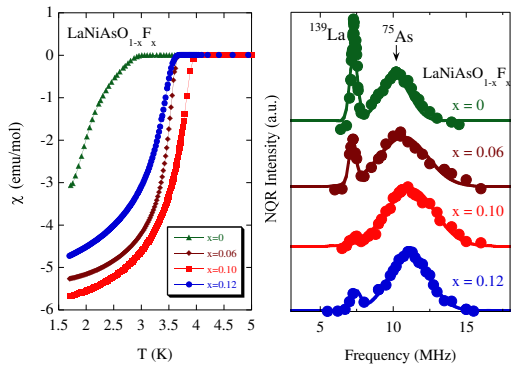}% Here is how to import EPS art
\caption{\label{fig1} (color online)
(left) DC susceptibility below $T$=5 K. (right) NQR spectra at  $T$ = 4.2 K.
% in LaNiAsO$_{1-x}$F$_x$ ($x$=0, 0.06, 0.10, 0.12). 
The baseline (the horizontal line associated with each spectrum) is shifted for clarity. The curves are guides to the eyes.
The left peak is from the  $\pm7/2\leftrightarrow \pm5/2$ transition of $^{139}$La and the right peak is from $^{75}$As. Except for $x$=0, the relative intensity of $^{139}$La is under-represented since the spectrum was taken with a repetition time of RF pulses which is suitable for $^{75}$As but not long enough for $^{139}$La.   }
\end{figure}

Figure 1 shows the the DC susceptibility around $T_c$ and the NQR spectra of $^{75}$As ($I$=3/2). The nuclear quadrupole  frequency $\nu _{\rm Q}$ increases slightly with increasing F-content.    
\begin{figure}[h]
\includegraphics[width=7.5cm]{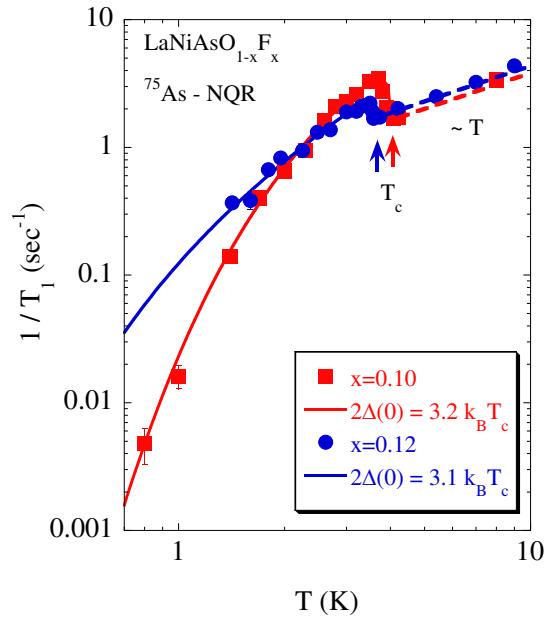}% Here is how to import EPS art
\caption{\label{fig2} (color online) The $T$ dependence of the spin lattice relaxation rate, $1/T_1$, for LaNiAsO$_{1-x}$F$_x$ ($x$=0.10 and 0.12). The arrows indicate $T_c$. The broken straight lines show the relation $1/T_1 \propto T$, and the curves below $T_c$ are fits to the BCS model with the gap size indicated in the figure.}
\end{figure}
Figure 2 shows the $T$ dependence of the spin lattice relaxation rate, $1/T_1$, at zero magnetic field, for the two samples with higher $T_c$.  The $T_1$  was measured at the  peak of the spectrum and determined  from an excellent fit of the nuclear magnetization to the single exponential function 1-$M(t)$/$M_0$ = $\exp$(-3$t/T_1$), where $M_0$ and $M(t)$ are the nuclear magnetization in the thermal equilibrium and at a time $t$ after the saturating pulse, respectively. 

 As seen in the figure, 
$1/T_1$ shows a well-defined coherence (Hebel-Slichter) peak just below $T_{\rm c}$, which is a characteristic of  superconductors with an isotropic gap.  This is in sharp contrast to various Fe-pnictides reported so far \cite{Matano,Ishida,Grafe,Kawasaki,Fukazawa,Matano2}. At low temperatures, $1/T_1$ decreases as an exponential function of $T$. 
 The solid curves in Fig. 2 are calculations using the BCS model. The $1/T_{1S}$ in the superconducting state is expressed as
%\begin{eqnarray}
$\frac{T_{1N}}{T_{1S}} = \frac{2}{k_BT}\int\int (1+\frac{\Delta^2}{EE'})N_s(E)N_s(E')f(E)[1-f(E')]\delta(E-E')dEdE'$
%\end{eqnarray}
where $1/T_{1N}$ is the relaxation rate in the normal state, $N_s(E)$ is the superconducting density of states (DOS),  $f(E)$ is the Fermi distribution function and $C =1+\frac{\Delta^2}{EE'}$   is  the "coherence factor". 
   Following Hebel \cite{Hebel}, we convolute  $N_{s}(E)$ with a broadening function $B(E)$ which is  approximated with a rectangular function centered at $E$ with a height of $1/2\delta$. The solid curves below $T_c$ for the two samples shown in Fig. 2 are  calculations with 2$\Delta(0)=3.2 k_BT_c$ and $r\equiv\Delta(0)/\delta$=5 for LaNiAsO$_{0.90}$F$_{0.10}$, and 2$\Delta(0)=3.1 k_BT_c$ and $r$=1.5 for LaNiAsO$_{0.88}$F$_{0.12}$.  

Such $T$-dependence of $1/T_1$ in the superconducting state is in striking contrast to that for  Fe-pnictides where no coherence peak was observed and the $T$-dependence at low-$T$ does not show an  exponential behavior. 
The striking difference 
%in the $T$-variation of $1/T_1$ of Fe-pnictides and Ni-pnictide 
may be ascribed to the different topology of the Fermi surfaces. 
For Fe-pnictides, it has been proposed that $d$-wave \cite{Graser,Kuroki2} or sign reversal  $s$-wave gap \cite{Mazin,Kuroki} can be stabilized  due to nesting by  the connecting wave vector $Q=(\pi, 0)$. 
%In principle, such gap should show a coherence peak just below $T_c$. However, it was shown that  magnetic scattering and/or impurity scattering between the two different bands can reduce the coherence peak just below $T_{\rm c}$ \cite{Chubukov,Parker,Hu,Bang,Nagai}. The impurity scattering can also alter the otherwise would-be-seen exponential $T$-variation at low temperatures. 
%
In LaNiAsO$_{1-x}$F$_{x}$, however, there is no such Fermi surface nesting \cite{Xu}, and thus the mechanism for the proposed gap symmetry does not exist.
Our result therefore highlights the important role of the Fermi-surface topology in the superconductivity of Fe-pnictides.

\begin{figure}[h]
\includegraphics[width=6.5cm]{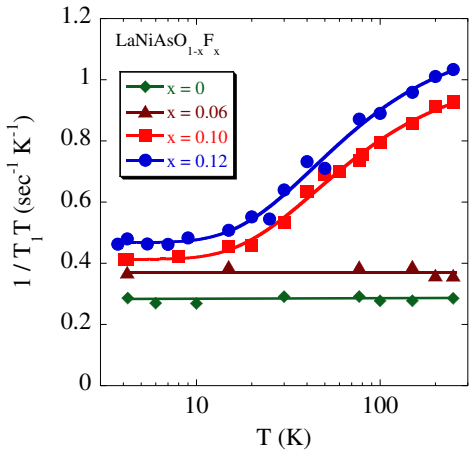}% Here is how to import EPS art
\caption{\label{fig3} (color online)
The quantity $^{75}$($1/T_1T$) in the normal state of LaNiAsO$_{1-x}$F$_x$ above $T_c$. The errors are within the size of the symbols. Below $T$=10 K, $1/T_1T$ is constant for all F-contents, indicating weak electron correlations. The straight lines are guides to the eyes. The curves for $x$=0.10 and 0.12 are fits to $1/T_1T=(1/T_1T)_0+b\cdot exp(-2E_g/k_BT)$ with $(1/T_1T)_0$=0.41 Sec$^{-1}$K$^{-1}$, $b$=0.61 Sec$^{-1}$K$^{-1}$ and $E_g/k_B$=22.5 K for $x$=0.10, and $(1/T_1T)_0$=0.46 Sec$^{-1}$K$^{-1}$, $b$=0.67 Sec$^{-1}$K$^{-1}$ and $E_g/k_B$=22 K for $x$=0.12}
\end{figure}

Figure 3 shows the quantity  $1/T_1T$ as a function of $T$. For $x$=0 and 0.06, $1/T_1T$ is a constant independent of $T$, which indicates that the electron correlations are weak as in conventional metals. This feature is also in striking contrast to the Fe analog. In LaFeAsO$_{1-x}$F$_x$, a magnetically ordered state is realized below $T_N$=140 K for $x$=0 \cite{Dai}. For small $x$, $1/T_1T$ increases with decreasing $T$ \cite{Kawasaki}, indicating electron correlations as seen in high-$T_c$ cuprates. Among others, the explanation based on  Fermi surface nesting is a promising scenario to account for the magnetic order \cite{Dong,Mazin}. Fermi surface nesting is also proposed to be responsible for  the spin fluctuations with $Q$=($\pi$, 0) \cite{Mazin,Kuroki}. In LaNiAsO$_{1-x}$F$_x$, however, such Fermi surface nesting does not exist, therefore  the spin fluctuations are not expected. Thus, the striking difference in the normal state between LaNiAsO$_{1-x}$F$_x$ and LaFeAsO$_{1-x}$F$_x$ can also be understood by the different topology of the Fermi surfaces.

\begin{figure}[h]
\includegraphics[width=7.5cm]{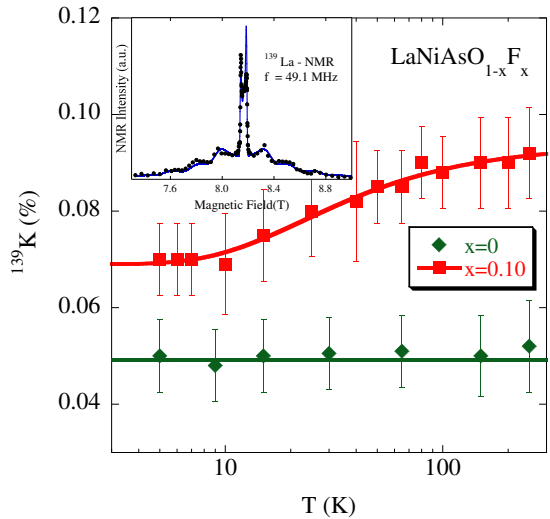}% Here is how to import EPS art
\caption{\label{fig4} (color online)
The $^{139}$La Knight shift in the $ab$-plane of LaNiAsO$_{1-x}$F$_{x}$ ($x$=0 and 0.10) above $T_c$. The curve for $x$=0.10 is a fit to $K=K_0+K_1\cdot exp(-E_g/k_BT)$ with $K_0$=0.07\%, $K_1$=0.025\% and $E_g/k_B$=23 K. The straight line for $x$=0 is a guide to the eyes. The inset shows the NMR  spectrum at 49.1 MHz and $T$=5 K for $x$=0.10. It consists of two central peaks corresponding to  respectively $\theta$=90$^o$ (left) and 41.7$^o$ (right),
accompanied by six satellite peaks since the nuclear spin is $7/2$. Here $\theta$ is the angle between $H$ and the $c$-axis. The curve is a theoretical calculation with $\nu_Q$=2.35 MHz for a ramdomly-oriented powders. The sample appears to be partially-oriented in the $ab$-plane, as in LaFeAsO$_{0.9}$F$_{0.1}$ \cite{Grafe} and PrFeAsO$_{0.89}$F$_{0.11}$ \cite{Matano}.  }
\end{figure}

On the other hand,  for LaNiAsO$_{0.90}$F$_{0.10}$ and LaNiAsO$_{0.88}$F$_{0.12}$, $1/T_1T$ increases with increasing $T$. A similar feature was also seen in highly-doped LaFeAsO$_{1-x}$F$_x$ with $x\geq$0.1 \cite{Ishida,Ahilan,Oka} and Ba(Fe$_{0.9}$Co$_{0.1}$)$_2$As$_2$ \cite{Ning}. Clearly, such behavior cannot be ascribed to electron correlations which are believed to be responsible for a similar phenomenon in high-$T_c$ cuprates (called pseudogap) \cite{Timsk}. It was proposed that such pseudogap behavior in LaFeAsO$_{1-x}$F$_x$ is   due to the band structure \cite{Ikeda}. In LaFeAsO$_{1-x}$F$_x$ with larger electron doping of $x\geq$0.1, the  large  DOS due to the hole pocket around the $\Gamma^{'}$ point, namely ($\pi$, $\pi$), sinks to  below the Fermi level  \cite{Ikeda}. At low $T$, this part of DOS does not contribute to $1/T_1$. Upon increasing $T$, however, the thermal activation associated with such  DOS 
%(thermal broadening of the energy level of the hole pocket) 
will contribute to $1/T_1$, giving rise to the pseudogap behavior. At present, the origin for  the pseudogap behavior in LaNiAsO$_{1-x}$F$_x$ is unclear. But there is a good possibility that the same mechanism \cite{Ikeda} also applies.  Upon doping, the bands around X and R points in LaNiAsO$_{1-x}$F$_x$ \cite{Xu} could sink to below the Fermi level \cite{Fang}, resulting in a  pseudogap behavior similar to LaFeAsO$_{1-x}$F$_x$ ($x\geq$ 0.1).
The $^{139}$La Knight shift ($K$) shown in Fig. 4 supports such speculation, since the spin part of $K$ is proportional to the DOS at the Fermi level ($N(E_F)$), while $1/T_1T \propto N(E_F)^2$.  
The curve for $x$=0.10 and 0.12 in Fig. 3 are fits to $1/T_1T=(1/T_1T)_0+b\cdot exp(-2E_g/k_BT)$. The first term is proportional to $N(E_F)^2$ at low $T$, and the second term comes from the thermal activation effect of the DOS beneath the Fermi level. The  obtained $E_g/k_B$ is 22.5 K and 22 K for $x$=0.10 and 0.12, respectively. The curve in Fig. 4 is a fit to $K=K_0+K_1\cdot exp(-E_g/k_BT)$ with resulting $E_g/k_B$=23 K, in good agreement with the $1/T_1T$ result.

\begin{figure}[h]
\includegraphics[width=6cm]{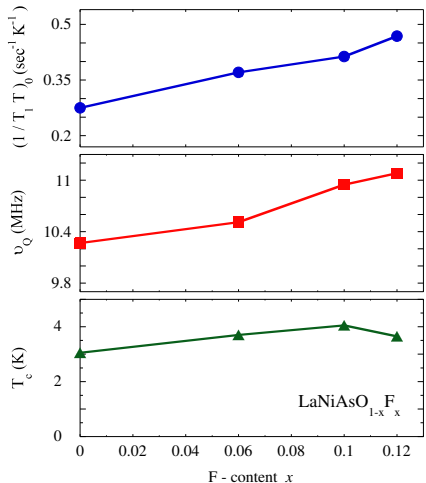}% Here is how to import EPS art
\caption{\label{fig5} (color online) Phase diagram obtained from the present study. $\nu_Q$ is the NQR frequency at $T$=4.2 K, and $(1/T_1T)_{0}$ is the averaged value for $T_c\leq T\leq$ 10 K, which is proportional to the squared DOS at the Fermi level. }
\end{figure}

Figure 5 summarizes the results. Upon replacing O with F, the value of $(1/T_1T)_{0}$ at low temperatures increases, indicating electron doping into the system. This is also supported by the F-content dependence of $\nu_Q$ (middle panel) which is generally dominated by the on-site charge distribution. The $T_c$  initially increases upon doping, but becomes less $x$ dependent beyond 0.06 \cite{Li}. And most remarkably different, $T_c$ is smaller by nearly an order of magnitude compared to LaFeAsO$_{1-x}$F$_x$. If the superconductivity in LaFeAsO$_{1-x}$F$_x$ is promoted by the spin fluctuations with $Q$=($\pi$, 0) as proposed \cite{Mazin,Kuroki}, then the much lower $T_c$ in  LaNiAsO$_{1-x}$F$_x$ can be naturally understood, since such spin fluctuation is not expected due to the different Fermi surface topology and indeed is not observed experimentally.

In summary, we have presented the  NQR results on  the electron-doped, low-$T_c$ nickel-pnictides   LaNiAsO$_{1-x}$F$_x$ ($x$ = 0, 0.06, 0.10 and 0.12). We find that the superconducting gap is  fully opened on the entire Fermi surface. Namely, the spin-lattice relaxation rate shows a well defined coherence peak just below $T_c$ and decays exponentially with further decreasing $T$. In the normal state, no antiferromagnetic spin fluctuations were observed. These features are in striking contrast with LaFeAsO$_{1-x}$F$_x$ where high-$T_c$, unusual superconductivity emerges in a background of moderate  antiferromagnetic spin fluctuation. All these differences are understood by the different topology of the Fermi surface for the two classes of materials. Our results highlight the peculiarity and importance of the Fermi surface topology in LaFeAsO$_{1-x}$F$_x$ and shed new light on the mechanism of  superconductivity and electron correlations there.

%************************acknowledgments*************************************************************************************
We thank  Z. Fang, H. Ikeda, T. Kambe, H. Kontani, K. Kuroki and Z. Wang for useful discussion.
This work was supported in part by research grants from JSPS and MEXT (No.17052005 and No. 20244058). The works in CAS and RUC were supported by NSF of China. 

\vspace{0.2cm}
$*$ Electronic address: zheng@psun.phys.okayama-u.ac.jp
%1
%\bibliography{apssamp}

\end{document}